# Planning Security Services for IT Systems

September 2014, Version 1.04

Marie Henderson[1] and Howard Philip Page[2]

**Abstract.** Often the hardest job is to get business representatives to look at security as something that makes managing their risks and achieving their objectives easier, with security compliance as just part of that journey. This paper addresses that by making planning for security services a 'business tool'. Recognising that no single approach for employing security services will ever meet every need, the authors focus on a high-level approach with low initial and on-going resource cost. The result is a common basis for business representatives and security practitioners to discuss the 'what' and 'how' of protecting IT systems for the outyears. Business representatives are provided with a basis to assist in sourcing funding and the potential for grouping blocks of security work allows for saving on resource costs. During the development and refinement of the approach the authors unearthed other benefits that are also presented.

## 1. Introduction

The application of security services for an IT system has often been done in an ad-hoc way, either only in response to incidents or as part of system development. It is recognised, however, that as risks change over time, a more organised and consistent approach is needed to manage changes in risks to system operation and information. The approach here uses a so-called Plan of Security Services for an IT System (or Plan).

Risk owners themselves are usually reliant on security practitioners for advice concerning the security services they need to apply. At the same time organisational security practitioners rarely have budgets to cover anything beyond implementation of the organisation wide security programme. So communication between risk owners and security is normally required to agree the security services performed for IT systems.

The approach for organising, communicating and managing security services for IT systems described herein was developed initially in response to a need to include business owners in a process that results in describing an appropriate level of security services for a system. Original objectives were to collect relevant information regarding an IT system's security and use this to plan security services in a way that indicated gaps, facilitated communication between key parties, was simple to implement, could meet policy requirements and assisted budget and work planning. Other benefits uncovered through the implementation of the approach are also outlined below in the discussion section below.

Some related approaches were found through review of literature and general enquiries. Some organisations have reported the inclusion of security services as part of an IT system plan or roadmap. However these have not contained the context information or reference section as described here and is found in more risk mature sectors such as the mining and resource sector (where risk management has long been a primary business consideration).

---

[1] Corresponding author hndrsnmr@yahoo.com.au
[2] The views and conclusions herein are those of the authors and do not necessarily represent the views of any particular organisation. The authors are indebted to the New Zealand Department of Internal Affairs (Security and Risk Group), for releasing Information Assurance processes for use.

The approach has similarities to that presented in [1]. In [1] the author proposes the use of control groups from NIST SP 800-53 [2] for implementing a roadmap for organisational level security. The NIST SP 800-53 approach can be used for IT system security as well with the controls split across common controls, so called *system specific controls* or hybrid controls. See also NIST SP 800-18 [3] and the related guide [4] which focus on system security plans as distinct from a plan for security services for an IT system.

As stated above with existing approaches found in certain industry sectors, the context information gathered is not as extensive or in a centralised system specific form. These approaches also lack a plan for security services and the work required to get underway is initially larger. A Statement of Applicability against ISO 27001/2 [7]/[8], the NZISM (New Zealand Information Security Manual) [5] or its parent manual the Australian ISM (Information Security Manual) [6] could also be regarded as having similarities, but have similar limitations to those of using NIST SP 800-53.

No one approach for employing security services will ever meet every need. The intent here is to describe an approach with low initial and on-going resource costs and where engagement with the risk and business representatives is underway early and an objective of the approach. A focus of this approach is to facilitate co-operation and communication between security practitioners and business representatives. As is always the case, the security practitioner must determine the practicality of an approach to their situation and also weigh up the expected benefits and costs.

The approach herein describes *what to do* based on a template showing *how it can be done*. The template is divided into two main parts: one being the business and related risk context and reference documentation relevant to the system's security, the other being the systems security service plan which is built by the security practitioner based on the information from the first part. These two parts are discussed below with reference to the template in Appendix A. The benefits of the approach are then discussed.

## 2. Part 1 – Framing Security within the Business and Risk Context

The first part of the Plan covers the business and risk context information relevant to the IT systems security as defined by the risk owner and suitable business representatives. It also covers all the security documentation and reports relevant to the IT system. The intention is not to get fixated on the information-gathering phase but rather begin to build the picture with the important risk and business representatives to get the Plan underway. A rough guide is to spend a few hours (including interviews with the risk and/or business representatives) rather than a few days building the Plan.

- Section 1 – High Level Information: records the basic details of the IT system and stakeholders.
- Section 2 – Assumptions: lists assumptions relevant to the Plan. Examples are included such as the Plan not forming an agreement covering funding, scheduling etc. (but forms a basis for such to occur).
- Section 3 – Business Context Information Relevant to Security: gathers simple business context information affecting security requirements for the IT system used to inform the Plan and provide rationale. For example if system availability were critical for business operation this would imply that Business Continuity Planning (BCP) is an important area. In this section, under "business impact" the use of the organisation's risk criteria, as defined in the organisation's corporate Risk Management Framework indicates how important the business operations are that rely on the IT system. This approach is pragmatic and broadly consistent with standards like [9], [2] and [10], whilst having lower complexity. The method of evaluation used to determine and record business impact ratings should be consistent across the organisation as this aids communication.

- <u>Section 4 – System Documentation and Reports</u>: collects the relevant documentation and previous security assessment reports. This could form part of a broader asset management plan, but it can be useful to have a dedicated view of the security relevant documents. Note this can be a good starting point as the documents and reports can provide context information to fill in Part 1 – especially if risk assessments, system operation guides and other documents with relevant detail are available.

- <u>Section 5 – Other Details</u>: gathers pertinent information for setting the security context for the system that does not fit under headings above. For example system dependencies can be critical to consider when determining the system security context, but are often overlooked by business and/or risk representatives and not captured under documentation.

- <u>Section 6 – Acknowledgements</u>: is optional and used where clarity of the information needs to be formalised. Note however that acknowledgements are not the focus for the Plan and should not delay development or commencement of activities. The only bureaucratic barriers that really count are those that come with funding and scheduling. The idea here is to get to that point as rapidly as possible, not add any delays. So acknowledgements should not detract from the real focus points.

## 3. Part 2 – Security Service Planning and Schedule

Part 2 is where the security practitioner presents their overview of the required security services for discussion with the risk owner and business representatives. Section 6 of the template contains the table for security service planning. This example has listed Service Options based on the NZISM [5] with the addition of Privacy Impact Assessments. The listed Service Options may be based on any relevant published standard or an organisation's own listing.

The security practitioner uses the information from Part 1 to determine the frequency with which each security service should be performed along with an explanation of why this is the case under the Rationale. Note this includes requirements as dictated by applicable legislation, policy and standards as collected under Section 5. For our example the NZISM requires system certification every 2-3 years and so the example Plan coverage shown has been designed to achieve only this.

The table in Section 6 does not specifically identify projects, but rather focuses on significant changes to the system as a trigger for undertaking a particular security service, for example a significant software upgrade. The reason for this is a project itself may not trigger the requirement for a particular security service. The table is focused towards the on-going frequency of security services to be applied to a system in a manner that is agnostic of the System Development Lifecycle (SDLC) whilst noting that changes to an existing system can also trigger the need for some security services to be performed.

Section 7 of the template is then completed by constructing a schedule of the security services to be performed over a relevant period providing foresight for business planning, in this example a 2-3 year period to cover the 2-3 year system certification requirement of the NZISM. Guidelines for estimating hours are provided in a separate table. The schedule and hours should be estimates as the focus is not to spend too much time fine-tuning details, but rather to assist the conversation with the risk owner and business representatives and in turn their budget and work planning.

In this case projects are best broken down under their own schedule as for one thing a project budget is usually separate from the on-going operational budget. The production system schedule can take account of security services performed under projects. For example, a penetration test when performed on a system immediately after putting the project changes in to production, can defer the need for further production system penetration testing as per the frequency requirement given in Section 6. This however must be carefully monitored in terms of project delays or cancellations to ensure appropriate testing and assessment is performed in a timely way on the production system irrespective of any project changes.

As already noted above, but worth repeating, the preparation of a Plan should be in the order of a few hours not days. Once a Plan is prepared it should be used to engage with the risk owner, business representatives and other relevant parties to agree on the security services performed along with the funding and a schedule. This agreement may be formalised through existing mechanisms the organisation employs, such as a memorandum. The focus should be to get things underway, especially the conversations and relationships between security and business and risk representatives and some work from the Plan, not to have everything in a Plan agreed before any work commences. The Plan itself remains a living document and should be updated accordingly and utilised in the on-going engagement and communication between the parties.

## 4. Discussion

The approach herein of using a single Plan to gather the business and risk context information relevant to security for an IT system along with simple rationale and schedule for security services to be performed on an IT system has the following benefits.

1. Low initial and on-going maintenance costs in terms of time and resource requirements.
2. Provides a way to engage and communicate with key parties such as risk owners, business representatives, project managers and IT system managers.
3. Provides a basis for agreeing funding and scheduling of security services for an IT system.
4. Is flexible to meet the organisations requirements be it formal as with system certification requirements in the NZISM or informal requirements the organisation may define.
5. Is flexible to incorporate changes bought about through projects or other system or organisational changes.
6. Provides a useful source of security relevant references (context information, documentation and reports) for the IT system.
7. Provides a basis for reporting to risk owners and other business representatives and maintaining these communication channels over time.
8. Assists risk owners communicating with auditors, service customers and stakeholders that due diligence is been performed in terms of risk management.
9. Enables the collection of schedules from multiple Plans for security management to obtain an overview of IT systems security services to track progress, manage resourcing and to obtain discounts through scale by collecting work where external security service providers are utilised (see Appendix B).

Many of the points have been discussed already or are apparent. We specifically cover 4 and 7-9 in further detail here. Further to 4 above this approach can be used to meet a number of government requirements: in connection to the NZISM it can be an input to a System Security Risk Management Plan and used to assist with achieving system certification [5], with the New Zealand Government Web Toolkit Guidance on Security ad Privacy Management it contributes to aspects of the quality assurance framework [11] and the mandatory requirement INFOSEC6 of [12].

With 7 it has been noted that the Plan itself provides a basis for facilitating communication with risk owners and business representatives. It can also be used as part of on-going reporting to these stakeholders, for example as part of a larger collective specific IT system report or for high level reporting covering general IT system security.

With 8 as the information in the Plan itself is not overly sensitive, it may be more broadly circulated than individual security assessments and reports and used to provide assurance to auditors, service customers and stakeholders that security and risk are being well managed and assist with showing that the risk owner is performing due diligence. It is evidence that on-going risk management is being undertaken by appropriate parties.

The final point 9 notes that multiple Plan schedules can be collated as shown in the sample IT systems resourcing and progress plan overview of Appendix B. This overview assists with management of delivery and resourcing of the security services. It also assists with the grouping of blocks of work for obtaining better rates when external service providers are sourced. For example: the penetration testing and vulnerability assessment for Project 1-IT system A, IT system C and Project 3-IT system D in the April-July 2014 period may be put out for bids collectively, saving resource time and encouraging vendors to return competitive bids to secure the work.

Finally note that although simple tables have been used in the examples provided in appendices, the approach can be implemented in this way or using management applications or tools already in place within the organisation or IT business. The tables used here are the minimum required to demonstrate the approach.

## 5. References

1. B. Kulkarni, A Sustainable and Efficient Way to Meet Clients' Growing Security Expectations – Achieving Holistic Security with NIST SP 800-53, *ISACA Journal* **6**, 42-45 (2013).

2. U. S. Department of Commerce – Joint Task Force Transformation Initiative, Security and Privacy Controls for Federal Information Systems and Organizations, *National Institute of Standards and Technology Special Publication 800-53 (NIST SP 800-53)*, Revision 4, April (2013). csrc.nist.gov

3. U. S. Department of Commerce and National Institute of Standards and Technology, Guide for Developing Security Plans for Federal Information Systems, *National Institute of Standards and Technology Special Publication 800-18 (NIST SP 800-18)*, Revision 1, February (2006). csrc.nist.gov

4. SANS Institute, System Security Plan Development Assistance Guide, V2.0 (2003). http://www.sans.org/projects/systemsecurity.php

5. Government Communications Security Bureau, New Zealand Information Security Manual, v1.01 June (2011). http://www.gcsb.govt.nz/

6. Australian Signals Directorate, Information Security Manual – Controls (2014). http://www.asd.gov.au/

7. International Organization for Standardization/International Electrotechnical Commission 27001:2013, *Information Technology – Security Techniques – Information Security Management Systems – Requirements*. http://www.iso.org/

8. International Organization for Standardization/International Electrotechnical Commission 27002:2013, *Information Technology – Security Techniques – Code of Practice for Information Security Controls*. http://www.iso.org/

9. U. S. Department of Commerce and National Institute of Standards and Technology, *Standards for Security Categorization of Federal Information and Information Systems, Federal Information Processing Standards Publications FIPS PUB 199*, February (2004). csrc.nist.gov

10. Attorney-General's Department, Protective Security Governance Guidelines - Business Impact Levels, *Protective Security Policy Framework*, 21 June (2011). www.protectivesecurity.gov.au

11. New Zealand Government, Web Toolkit - Security and Privacy Assurance, 21 May (2014). webtoolkit.govt.nz

12. Attorney-General's Department, Securing Government Business – Protective Security Guidance for Executives, *Protective Security Policy Framework*, May (2012). http://www.protectivesecurity.gov.au/

# Appendix A – Sample of Security Services for an IT System

## 1 High Level Information

| | |
|---|---|
| **System name** | *The name of the IT system covered in the Plan* |
| **System risk owner** | *The name and position of the risk owner for the business operations that the IT system supports* |
| **System service owner** | *The names and position of the business managers for the main business operation or services that the IT system supports* |
| **System IT operation manager** | *The name and position of the IT operational manager for the IT system* |
| **Project X manager** | *The name and position of the project manager for Project X* |
| **System users** | *Brief description of the business users of the IT system (or related services)* |
| **Other stakeholders** | *List of any other business risk stakeholders* |
| **Document information** | *The name and position of the preparer of the Plan and the date it was prepared (to include detail to define versions)* |

## 2 Assumptions

The following assumptions relate to this Plan:

- This Plan does not take account of costs, scheduling, or prioritisation for the security services proposed but rather provides a basis for such considerations to be undertaken
- The risk owner retains ultimate accountability and responsibility for risk decisions relating to the business operations and the sourcing of IT system support for those business operations
- *Any other assumptions specific to the Plan for the particular IT system.*

## 3 Business Context Information Relevant to Security

| Business Aspect | Details | | | |
|---|---|---|---|---|
| **Business description and objectives as related to the IT system** | *A 'set the scene' description of the business and how the IT system supports these along with the details of specific business objectives that the IT system must meet.* | | | |
| **Business Impact** | *The business impact ratings relevant to the organisations corporate risk criteria* | | | |
| **CIAP (Critical, Highly Important, Important, Some Importance, Unimportant, Not Applicable)** | Confidentiality: *List ratings as judged by business risk owner.* | Integrity: | Availability: | Privacy: |
| | The most important requirement in terms of information protection is: *information protection requirement as judged by the business risk owner.* | | | |
| **Information classification** | *The classification of the information that resides on or is transferred by the IT system as per the organisations information classification policy.* | | | |
| **Planned system development activities** | Next 6 months | Next 6-12 months | Next 12-24 months | Next 24-36 months |
| | *List any planned items.* | | | |
| Would changes in the business context affect the IT systems security profile? | | | | *Yes/No* |

## 4 System Documentation and Relevant Reports

| Document/Report Type (include document and report titles where relevant) | Document and Report Location/Link | Date | Verified |
|---|---|---|---|
| Existing Plan of Security Services for the IT system: | *Document file number and/or link* | *Month and year* | Y/N |
| Number of security incidents or privacy breaches in previous 12 months (include direction to reports): | | | |
| System Operating Guide: | | | |
| System Security Plan | | | |
| BCP and Disaster Recovery Plan (include direction to test results): | | | |
| Risk Assessments: | | | |
| Privacy Impact Assessments: | | | |
| Design Reviews: | | | |
| Penetration Tests: | | | |
| Vulnerability Assessments: | | | |
| Code Reviews: | | | |
| Certifications: | | | |
| IT General Control Assessments: | | | |
| Physical Security Assessments: | | | |
| Other (as applicable to the IT system security): | | | |

## 5 Other Details

| | | |
|---|---|---|
| **Dependencies** | The other systems this system is reliant on:<br><br>The other system that rely on this system: | |
| **Rationale for employing security services** | The risk owner requires positive assurance? | Y/N |
| | System reports directly to stated organisational outcomes (e.g. features in Statement of Intent, Annual Report etc.)? | Y/N |
| | Disaster Recovery/resilience measures have been tested in the last 12 months? | Y/N |
| | System contains or transfers information with a security classification? | Y/N |
| | System has automated transaction interfaces with external entity systems or is a point of information aggregation? | Y/N |
| | System is internet facing? | Y/N |
| | Other drivers (explain) | |
| **Top 3 Future Risks** | *List top three business risks to operation from the risk owners point of view.* | |
| **Applicable Legislation, Policy and Standards** | *List the primary applicable legislation, policy and standards the IT system must comply with.* | |
| **Control Landscape** | *Record any relevant control description or categorisation. (These may be only available in a high level form through to having detailed control documentation.) For example it may be helpful to record that only detective and compensating controls are used and that there is an absence of preventative controls and why this is so.* | |

# 6 Acknowledgements

| Name and Position | Signature | Date |
|---|---|---|
| *Name and position of risk owner* | As risk owner of the business operations utilising this IT System I have reviewed and confirm the information given above with respect to my business risks: | |
| *Name and position of IT security head* | As the IT security head I have reviewed and confirm the information as given above as relevant to IT security for this IT system: | |

# 7  Security Service Planning for IT System

| Service Area | Service Option | Frequency | Rationale[3] |
| --- | --- | --- | --- |
| **Information Security Risk Management** | Information Classification Evaluation | Where required | To occur whenever significant changes to the information assets held, transferred or processed are made. |
| | IT Risk Assessment | Every 2 years or when significant changes occur | A risk assessment for the IT System A (covering relevant aspects of Project 1) was performed in February 2015. A risk assessment must be performed every two years as part of the system certification and accreditation. |
| | Privacy Impact Assessment | Where design changes require | A Privacy Impact Assessment should be conducted when design changes with material impacts to privacy protections are planned. Refer also Security Design/Architecture review. |
| | Vulnerability Assessment | Every year or when significant changes occur | To occur annually with penetration testing to review the status of vulnerabilities associated with the system or where significant changes to the system occur. |
| | Penetration Testing | Every year of when significant changes occur | To be performed annually or when significant changes to the system occur that may introduce new vulnerabilities. |
| | Code Review | Where significant changes are made to the system code | To occur whenever significant changes to the system code are made (as with Project 1). |
| | Security Incident Management | Every 3 months | Regular security incident reporting to be provided. Planning to include provision for this and an occurrence of 1-2 incidents a year. |
| | Physical Security Assessment | Every 2 years or when significant changes occur | The physical security of the hosting data centre must be reviewed every 2 years or when significant site changes occur. A physical security review was performed in January 2015. |
| **Security Compliance** | Security Design/Architecture Review | Where design changes require | To occur whenever significant changes to the system design are made (as with Project 1). |
| | System Audit | Every 2 years or when significant changes occur | A system audit must be performed as part of system certification or where significant changes are made. |
| | Compliance Documentation | Every 2 years or when significant changes occur | Certification report and related documentation to be created as part of system certification – refer Certification below. |
| | Certification | Every 2 years or when significant changes occur | Certification is required to be performed every 2-3 years to meet compliance requirements. Certification is also triggered whenever significant changes to the IT system are made. |
| | Functional Testing (Security) | Where design changes require<br>Where incident management identifies weaknesses | No special requirement. |
| **Security Governance** | Security Governance Framework | When changes require | Planned to occur as part of next certification cycle to cover recent organisational and system security governance and external audit and reporting requirements. |
| | Policy and Standards | When changes require | No special requirement. |
| | Security Awareness Training | When changes require | No special requirement. |
| | Business Continuity Planning | At least one exercise a year. | BCP and IT Service Continuity is in place but newly implemented so a comprehensive exercise is planned and followed by annual test and reviews to ensure availability requirements are met. |

---

[3] For the purpose of the example we loosely used a fictitious system that: is subject to NZISM; operates within a changing organisational environment; has defined availability requirement; contains some custom code; holds some private data; and requires some protection from hostile networks to which it connects.

## 8 Security Service Delivery Schedule for IT System A with Project 1

| | | Schedule Summary | | | | | | | | | | | | | | | | | | | | | | | | | | | | |
|---|---|---|---|---|---|---|---|---|---|---|---|---|---|---|---|---|---|---|---|---|---|---|---|---|---|---|---|---|---|---|
| | | 2015 | | | | | | | | | | | | 2016 | | | | | | | | | | | | 2017 | | | | | |
| | | JAN | FEB | MAR | APR | MAY | JUN | JUL | AUG | SEP | OCT | NOV | DEC | JAN | FEB | MAR | APR | MAY | JUN | JUL | AUG | SEP | OCT | NOV | DEC | JAN | FEB | MAR | APR | MAY | JUN |
| **Production system schedule** | | PSA, ICE | RA | SIM | | SIM | | BCP, SA | SGF, CD | SIM, SC | | | SIM | | SIM | | | SIM | | BCP | | SIM | | | SIM | PSA | RA | SIM | | | SIM |
| **Estimate hours** | | 40 | 60 | 4 | | 4 | | 20+60 | 20+20 | 4+20 | | | 4 | | 4 | | | 4 | | 10 | | 4 | | | 4 | 40 | 40 | 4 | | | 4 |
| **Total hours by month** | | 40 | 60 | 4 | | 4 | | 80 | 40 | 24 | | | 4 | | 4 | | | 4 | | 10 | | 4 | | | 4 | 40 | 40 | 4 | | | 4 |
| **Total hours by budget year** | | 108 | | | | | | | | | | | 156 | | | | | | | | | | | | 106 | | | | | | |
| **Project 1 schedule** | | | | SDR | | CR | PT, VA | PT, VA | | | | | | | | | | | | | | | | | | | | | | | |
| **Estimate hours** | | | | 40 | | 40 | 40+10 | 20+10 | | | | | | | | | | | | | | | | | | | | | | | |
| **Total hours by month** | | | | 40 | | 40 | 50 | 30 | | | | | | | | | | | | | | | | | | | | | | | |
| **Total hours by budget year** | | 160 | | | | | | | | | | | - | | | | | | | | | | | | - | | | | | | |
| **Overall budget year total hours** | | 268 | | | | | | | | | | | 156 | | | | | | | | | | | | 125 | | | | | | |

| Abbreviation for Service Option | | Guidelines for Estimating Sum Total Hours for all Participants |
|---|---|---|
| **ICE** | Information Classification Evaluation | 10 hours for a stand alone system up to 20 hours for a complex system |
| **RA** | IT Risk Assessment | 40 hours for a stand alone system up to 160 hours for a complex system |
| **PIA** | Privacy Impact Assessment | 40 hours for a stand alone system up to 160 hours for a complex system |
| **VA** | Vulnerability Assessment | 10 hours for a stand alone system up to 80 hours for a complex system |
| **PT** | Penetration Testing | 40 hours for a stand alone system up to 100 hours for a complex system |
| **CR** | Code Review | 40 hours for a stand alone system up to 100 hours for a complex system |
| **SIM** | Security Incident Management | Determined on a system by system basis |
| **PSA** | Physical Security Assessment | 20 hours for a stand alone system up to 80 hours for a complex system |
| **SDR** | Security Design/Architecture Review | 20 hours for a stand alone system up to 80 hours for a complex system |
| **SA** | System Audit | 40 hours for a stand alone system up to 200 hours for a complex system |
| **CD** | Compliance Documentation | 40 hours and up to 80 hours depending on certification |
| **SC** | Security Certification | 20 hours for a stand alone system up to 80 hours for a complex system |
| **FTS** | Functional Testing (Security) | 40 hours for a stand alone system up to 160 hours for a complex system |
| **SGF** | Security Governance Framework | Determined on a system by system basis |
| **P&S** | Policy and Standards | 20 hours for a stand alone system up to 80 hours for a complex system |
| **SAT** | Security Awareness Training | 20 hours for a stand alone system up to 80 hours for a complex system |
| **BCP** | Business Continuity Planning | 20 hours for a stand alone system up to 80 hours for a complex system |

**Appendix B – Sample IT Systems Resourcing and Progress Plan Overview**

| IT System Name (Including Related Projects) | Security Resource Lead | Schedule Summary |||||||||||||||||||||||||||||||
|---|---|---|---|---|---|---|---|---|---|---|---|---|---|---|---|---|---|---|---|---|---|---|---|---|---|---|---|---|---|---|---|
| | | 2015 |||||||||||| 2016 |||||||||||| 2017 ||||||
| | | JAN | FEB | MAR | APR | MAY | JUN | JUL | AUG | SEP | OCT | NOV | DEC | JAN | FEB | MAR | APR | MAY | JUN | JUL | AUG | SEP | OCT | NOV | DEC | JAN | FEB | MAR | APR | MAY | JUN |
| **IT System A** | A. Bloggs | PSA | RA | SIM | | | SIM | BCP | SGF | SC | | | SIM | | | SIM | | SIM | BCP | | SIM | | | | SIM | PSA | RA | SIM | | | SIM |
| | | ICE | | | | | | SA | CD | SIM | | | | | | | | | | | | | | | | | | | | | |
| | | | | | | | | | | | | | | | | | | | | | | | | | | | | | | | |
| **Project 1** | | | SDR | | CR | PT | PT | | | | | | | | | | | | | | | | | | | | | | | | |
| | | | | | | VA | VA | | | | | | | | | | | | | | | | | | | | | | | | |
| **Project 2** | | | | | | | | | | SDR | | CR | PT | PT | | | | | | | | | | | | | | | | | |
| | | | | | | | | | | PIA | | | VA | VA | | | | | | | | | | | | | | | | | |
| **IT System B (external system)** | B. Bloggs | | | SIM | | | SIM | | | SIM | SC | | SIM | | | SIM | PSA | | SIM | | | SIM | | | SIM | | | SIM | | | SIM |
| **IT System C** | C. Bloggs | PSA | | BCP | RA | PT | | SA | SC | | | | | | | BCP | | PT | | | | | | | | PSA | | BCP | | | RA |
| | | | | | | VA | | CD | | | | | | | | | | VA | | | | | | | | | | | | | |
| **IT System D** | B. Bloggs | PSA | | | | BCP | | SA | CD | SC | | | | | | | PT | BCP | | | RA | | | | | PSA | | | PT | BCP | |
| **Project 3** | | | | RA | PT | | | | | | | | | | | | | | | | | | | | | | | | | | |

| Key | |
|---|---|
| Completed | |
| Funded | |
| Proposed | |